\documentstyle[aps,preprint,epsf]{revtex}
\newcommand{\beq}{\begin{equation}}
\newcommand{\eeq}{\end{equation}}
\begin{document}
\tightenlines

\title{Coulomb Energy of Nuclei}

\author{V.R. Shaginyan
\footnote{E--mail: vrshag@thd.pnpi.spb.ru}}
\address{
Petersburg Nuclear Physics Institute, Gatchina 188350, Russia}
\maketitle

\begin{abstract}
The density functional determining the Coulomb energy of nuclei is
calculated to the first order in $e^2$. It is shown that the Coulomb
energy includes three terms: the Hartree energy; the Fock energy; and
the correlation Coulomb energy (CCE), which contributes
considerably to the surface energy, the mass difference between
mirror nuclei, and the single-particle spectrum. A CCE-based
mechanism of a systematic shift of the single-particle spectrum is
proposed. A dominant contribution to the CCE is shown to come from
the surface region of nuclei. The CCE effect on the calculated proton
drip line is examined, and the maximum charge $Z$ of nuclei near this
line is found to decrease by 2 or 3 units. The effect of Coulomb
interaction on the effective proton mass is analyzed.
\end{abstract}


\section{Introduction}
Relatively weak Coulomb interaction substantially
affects the properties of nuclei owing to its long-range
character. It is reliably established that the electrostatic
Coulomb energy of the distributed charge is the
dominant term in the Coulomb energy of nuclei. This term,
also known as the Hartree energy, is proportional to
$Z^2e^2/R$, where $Z$ is the number of intranuclear protons,
$R$ is the nuclear radius, and $e$ is the proton charge; at
large values of $Z$, it induces the breakup of nuclei. At
the same time, there are a few less significant
contributions to the Coulomb energy. These include the Fock
exchange contribution and a number of terms
associated with the correlated motion of nucleons. It will be
shown below that, in constructing the density functional determining
the Coulomb energy of a nucleus, it is sufficient to calculate all
these terms to the first order in $e^2$. The presence of these terms
is clearly illustrated by the well-known Nolen--Schiffer anomaly
\cite{n}, which was discovered more than 30 years ago. This anomaly
was deduced from a comparison of the calculated mass
difference between two mirror nuclei with its experimental value.
The comparison revealed that, on average, the results of the
calculations performed without taking consistently into account the
correlation Coulomb energy (CCE) fall short of relevant
experimental values by 10\% \cite{n1,n2}. This discrepancy can be
removed by introducing charge-dependent forces \cite{csb} whose
strength is assumed to be preset by this discrepancy.  However, this
is possible if we are sure that nuclear dynamics is taken properly
into account and that all the Coulomb contributions are consistently
included. A new mechanism that enhances the contribution of Coulomb
interaction to the energy of the nuclear ground state and which is
caused by the presence of a surface (a general property of
equilibrium finite Fermi systems) was found in \cite{vs,bs}.
This contribution, which is proportional to the nuclear surface
$(Z^{2/3})$, made it possible to explain a dominant part of the above
anomaly.

The main objective of this study is to calculate consistently the
density functional determining the Coulomb energy. As the result of
this calculation, which will be performed to the first order in
$e^2$ and which will take into account the mechanism enhancing the
CCE contribution, we will be able to clarify the effect of this
mechanism on the single-particle spectrum of nuclei,
on the effective nucleon mass, and on the position of
the proton drip line. In what follows, we do not
highlight the difference between this functional and the
Coulomb energy unless this leads to confusion.

The ensuing exposition is organized as follows. A
general formulation of our approach to calculating the
Coulomb energy is given in Section 2. Section 3 is
devoted to calculating the systematic CCE-induced
shifts of the single-particle spectrum and the proton
drip line. The effect of Coulomb interaction on the
effective proton mass is considered in Section 4. Section 5
summarizes the basic results.

\section { Coulomb correlation energy}

In order to formulate the aforementioned mechanism and to consider
its effect, we will make use of the Hartree--Fock method employing
effective forces, which was substantiated within density-functional
theory \cite{rs}.
In this case, the energy $E$ of the nuclear ground state
has the form
\beq E=F_0[\rho_p,\rho_n] +F_c[\rho_p,\rho_n].\eeq
where $F_0$ is that part of the functional which is
due to nuclear forces. Since we assume that the charge symmetry of
nuclear forces takes place, $F_0$ depends identically on the
single-particle densities $\rho_p$ and $\rho_n$ of protons and
neutrons (these densities are determined by minimizing $E$). The term
$F_c$ is due to other interactions, which are weaker than nuclear
forces. The smallness of these interactions can be characterized by
the ratio a of the potential energy of these interactions between two
nucleons to the Fermi energy $\varepsilon_F\simeq 40$ MeV. For the
Coulomb interaction, this ratio is $\alpha=e^2/r_0\varepsilon_F
\simeq 0.03$, where $r_0$ is the mean spacing between nucleons in
equilibrium nuclear matter.  Therefore, it is sufficient to calculate
the functional $F_c$ in the first order in the relevant coupling
constant. In the case of Coulomb interaction, which is the main
subject of the present study, the functional $F_c$ will be calculated
to the first order in $e_2$.  This functional could be approximated
by some simple expression, as was done, for example, for the
functional $F_0$. However, experience gained in calculating the mass
differences between mirror nuclei (these mass differences are
directly determined by the functional $F_c$) proves the inefficiency
of such attempts.  Moreover, this oversimplified approximation does
not reveal the physics behind the functional $F_c$. Therefore, it is
preferable to calculate the functional $F_c$. In doing this, it is
assumed that the functional $F_0$ is known. It can be taken
in the form specified by the Skyrme interaction \cite{vb} or in
the form proposed in \cite{fa}. The latter is advantageous in
that it is characterized by a high precision, is simple in
use, and involves the bare nucleon mass. By construction, the
functional $F_0$ must contain the unnrenormalized mass $M$ of the
proton (neutron) rather than its effective mass $M^*$ because the
effective mass, as well as the nuclear spectrum, cannot be directly
included in it \cite{ksh}.  The single-particle distributions of the
proton and neutron densities ($\rho_p$ and $\rho_n$, respectively)
are specified as
\beq \rho_p({\bf r})=\sum_l n_p^l |\phi_p^l({\bf
r})|^2;\,\,\, \rho_n({\bf r})=\sum_l n_n^l |\phi_n^l({\bf r})|^2,
\eeq
where $n_p^l$ and $n_n^l$ are the occupation numbers for
single-particle proton and neutron levels, respectively, while
$\phi_p^l$ and $\phi_n^l$
are, respectively, the proton and neutron single-particle wave
functions determined from the Hartree--Fock equation \cite{vb}. The
term $F_c[\rho_p]$ is usually taken in the first order in the Coulomb
interaction in the Hartree--Fock approximation \cite{rs,vb}:
\beq F_c=\frac{e^2}{2}\left[\int \frac{\rho_p({\bf
r}_1)\rho_p({\bf r}_2)} {|{\bf r}_1-{\bf r}_2|}d{\bf r}_1d{\bf r}_2
-\int\left[\frac{\chi^0_p({\bf r}_1,{\bf r}_2,i\omega)
+2\pi\rho_p ({\bf r}_1)\delta({\bf r}_1-{\bf
r}_2)\delta(\omega)}{|{\bf r}_1-{\bf r}_2|}\right]
\frac{d{\bf r}_1d{\bf r}_2 d\omega}{2\pi}\right].\eeq
Here, $\chi^0_p({\bf r}_1,{\bf r}_2,\omega)$
is the linear response function for noninteracting
protons moving in the self-consistent single-particle field $V_p$.
The first and the second term on the right-hand side of Eq. (3) are,
respectively, the Hartree and the Fock density functional  ($F_c^H$
and $F_c^F$). The exchange term is usually approximated by the
expression \cite{vb}, \beq
F_c^F=-\frac{3e^2}{4}\left(\frac{3}{\pi}\right)^{1/3}
\int\rho_p^{4/3}({\bf r})d{\bf r}.
\eeq
However, the equality in (3) is not exact even in the first
order in the Coulomb interaction ($e^2$ order) because it
does not include the functional $F_c^{corr}[\rho_p({\bf r})]$ taking
into account the contribution to the Coulomb energy from the
correlated motion of protons under the effect of the effective
(residual) nuclear interaction
$R_{\alpha\beta}({\bf r}_1,{\bf r}_2)$. In the first order in $e^2$,
the CCE functional $F_c^{corr}$ has the form
\cite{bs}
\beq
F_c^{corr}[\rho_p]=-\frac{e^2}{2}\int
\left[\frac{\chi_{pp}({\bf r}_1,{\bf r}_2,i\omega)-
\chi^0_p({\bf r}_1,{\bf r}_2,i\omega)}
{|{\bf r}_1-{\bf r}_2|}\right]
\frac{d{\bf r}_1d{\bf r}_2 d\omega}{2\pi},
\eeq
where $\chi_{pp}({\bf r}_1,{\bf r}_2,\omega)$
is the exact function representing the linear response of
intranuclear protons to an external electric field.
It should be noted that the function $\chi_{pp}$
is completely determined by the functional $F_0$ \cite{bs} and
that it involves no smallness associated with the
Coulomb interaction. The total response function contains
three independent components --- $\chi_{pp}$, $\chi_{nn}$
and $\chi_{np}$ --- and satisfies the equation
\beq
\chi_{lm}=\chi^0_{l}\delta_{lm}
+\sum_k\chi^0_{l}R_{lk}\chi_{km}.\eeq
where integration
signs are omitted for the sake of simplicity.
The response function $\chi^0_{p}$ for noninteracting
nucleons (protons) moving in the self-consistent field
$V_p$ has the form
\beq \chi^0_p({\bf r}_1,{\bf r}_2,\omega)=\sum_{i,k}
n^i_p(1-n^k_p)\phi^{i*}_p({\bf r}_1)\phi^{i}_p({\bf r}_2)
\phi^{k*}_p({\bf r}_2)\phi^k_p({\bf r}_1)
\left[\frac{1}{\omega-\omega^{ik}_p+i\eta}
-\frac{1}{\omega+\omega^{ik}_p-i\eta}\right]. \eeq
The single-particle eigenvalues $\varepsilon^i_p$, the frequencies
$\omega^{ik}_p=\varepsilon^k_p-\varepsilon^i_p$,
the single-particle wave functions
$\phi^{i*}_p({\bf r})$, and the corresponding densities $\rho_p$ and
$\rho_n$ are determined by solving the single-particle
Hartree--Fock--like equations, or
the Kohn--Sham equations \cite{ksh},
\beq
\left(-\frac{\nabla^2}{2M}+V_p({\bf r})\right)\phi^i_p({\bf r})
=\varepsilon^i_p\phi^i_p({\bf r}), \eeq
which are derived by means of a conventional procedure that amounts
to varying Eq. (1) with respect to the single-particle functions
$\phi_{n,p}^l$ \cite{rs,vb}. The self-consistent
potential $V_{p}$ and the effective interaction $R_{\alpha\beta}$ are
given by
\beq
V_{p}({\bf r})=\frac{\delta(F_0[\rho_p,\rho_n]-T_0[\rho_p,\rho_n])}
{\delta\rho_p({\bf r})};\,\,\,
R_{\alpha\beta}({\bf r}_1,{\bf r}_2)=\frac
{\delta^2(F_0[\rho_p,\rho_n]-T_0[\rho_p,\rho_n])}
{\delta\rho_{\alpha}({\bf r}_1)\delta\rho_{\beta}({\bf r}_1)},\eeq
where $T_0[\rho_p,\rho_n]$ is the kinetic-energy
functional for noninteracting nucleons. Thus, we conclude that, to
the first order in $e^2$, the eventual form of the functional $F_c$ is
\beq
F_c=F_c^H+F_c^F+F_c^{corr}.\eeq
Equations (1), (3), (5), and (10) provide a basis for consistently
calculating the Coulomb energy and for taking into account the
Coulomb interaction effect on the properties of nuclei. The poles of
the response function c $\chi_{pp}$ determine the collective spectrum
of nuclei, while the residues at these poles govern the relevant
transition probabilities. As can be seen from Eq. (5), a dominant
contribution to $F_c^{corr}$ comes from collective isoscalar
surface vibrations whose excitation energies are much
lower than those of corresponding isovector modes.
Thus, the functional $F_c^{corr}$ is controlled by the isoscalar
components of the effective interaction.

It is instructive to consider the CCE for symmetric
infinite nuclear matter \cite{vs,bs}, in which case the relevant
equations are considerably simplified.
The function c $\chi_{pp}$
assumes the form
\beq
\chi_{pp}(q,\omega)=\frac{1}{2}\left[\frac{\chi^0(q,\omega)}
{1-R_{+}(q,\omega,\rho)\chi^0(q,\omega)}+
\frac{\chi^0(q,\omega)}
{1-R_{-}(q,\omega,\rho)\chi^0(q,\omega)}
\right],\eeq
and determines the functional $F^{corr}_c$:
\beq
F_c^{corr}[\rho_p]=-2\pi e^2\int\left[
\frac{\chi_{pp}(q,i\omega)-
\chi^0_p(qi\omega)}
{q^2}\right]\frac{d{\bf q}d\omega}{(2\pi)^4}.
\eeq
For symmetric matter, we have $\chi^0_p=\chi^0_n=\chi^0/2$; the
effective interaction $R(\rho)$ is similar to the local interaction
$f(\rho)$ between nucleons that was introduced by Migdal \cite{ab}:
\beq
\frac{p_F M}{\pi^2}R_{+}(\rho)=f(\rho);\,\,\,
\frac{p_F M}{\pi^2}R_{-}(\rho)=f^{\prime}(\rho).
\eeq
Considering that the isoscalar amplitude $f(\rho\to0)$ is
approximately equal to $-2.5$ and that $p_F M/\pi^2=\chi^0(0,i0)$,
one can see from Eq. (11) that, at relatively low densities
corresponding to the surface region of nuclei, the denominator
$1-R_{+}\chi^0$ vanishes at small values of $q$ and
$\omega$. Therefore, the function $\chi_{pp}$ develops poles on the
imaginary axis. These poles, which evince instability of
low-density nuclear matter \cite{ab}, lead to the divergence
of the integral in Eq. (12), and this prevents a
calculation of the CCE at these nuclear-matter densities. We
can conclude that the main contribution to this integral
comes from the isoscalar response function. From the
above, it also follows that the conventional procedure
based on calculating various quantities in the local
density approximation (LDA) in infinite uniform matter
fails in this case. In this connection, it is worth noting
that, although an attempt was made in \cite{be} to construct
the functional $F_0$ within the LDA, this is impossible for
the same reason.
It is convenient to calculate the CCE in semi-infinite
nuclear matter, where there is a region of low-density
nuclear matter near the surface. In the generalized
LDA, we then obtain \cite{vs,bs},
\beq F_c^{corr}[\rho_p]=\int\rho_p({\bf
r})e_c(\rho({\bf r}))d{\bf r},\eeq
where $e_c(\rho)$ is the CCE per proton and $\rho_p$ is the
single-particle proton density. The calculation of the energy
$e_c(\rho_p)$ for semi-infinite nuclear matter revealed that, at
the surface, this energy has a pronounced positive peak,
which corresponds to smoothing the divergence in (12)
for uniform nuclear matter \cite{bs}. It is convenient to
approximate the energy $e_c$ by the simple expression
\beq e_c(\rho_p(r))=
D\left[\frac{\rho_p(r)}{\rho_0}
\left(1-\frac{\rho_p(r)}{\rho_0}\right)\right]^{4/3},\eeq
where $2\rho_0=0.16\,fm^{-3}$ is the equilibrium nuclear density
and $D\simeq 6$ MeV. In contrast to the terms $F_c^H$ and $F_c^F$,
the energy $e_c(\rho_p)$, which has a pronounced peak at the nuclear
surface, makes a noticeable contribution  to the surface tension
$\sigma_c$,
\beq \sigma_c=\int_{-\infty}^{\infty}\rho_p(z)
[e_c(\rho_p(z))-e_c(\rho_p(-\infty))]dz.\eeq
Hence, the Weizs\"{o}cker mass formula must be supplemented with the
term $\Delta E$ whose contribution to the total binding energy of a
nucleus can be represented as \beq \Delta
E=a_1^{\prime}Z+a_2^{\prime}Z^{2/3}, \eeq
where $a_1^{\prime}\simeq -0.1$ MeV and $a_2^{\prime}\simeq 1.0$
MeV. Thus, we can see from Eq. (17) that the surface tension has an
isovector component. It will be shown below that this is so for the
effective mass as well. It follows from Eq. (17) that the surface
tension of the proton nuclear Fermi liquid is effectively greater
than that of neutron matter. This means that, when protons are added
to a nucleus, their binding energy becomes somewhat smaller because
of an increase in the surface of the proton liquid component. It
will also be proven below that this decrease in the proton binding
energy is sufficient for explaining the mass difference between
mirror nuclei. Further, the additional surface tension $\sigma_c$ and the
addition of extra neutrons are expected to induce opposite
modifications in the root-mean-square radius $<r_p>$ of the
intranuclear-proton distribution: the former would moderate an
increase associated with the latter. The preliminary calculations
reveal that this could explain the anomalously small increase in the
radius $<r_p>$ in going over from the $^{40}$Ca to the $^{48}$Ca
nucleus. The contribution $\sigma_c$ must also be taken into account
in considering fission barriers for heavy nuclei.

\section{Single-particle spectrum and the proton drip line}

Let us calculate the CCE-induced shift of the
single-particle proton excitation spectrum
$\varepsilon_p^l$. For this, we will use the well-known Landau
equation \cite{lan}
\beq
\frac{\delta E}{\delta n_p^l}=\varepsilon_p^l.\eeq
It follows from Eqs. (5) and (18) that this shift
$\Delta\varepsilon_p^l$ can be represented as
\beq
\Delta\varepsilon_p^l=\frac{\delta F_c^{corr}}{\delta n_p^l}
=-\frac{e^2}{2}\frac{\delta}{\delta n_p^l}\int
\left[\frac{\chi_{pp}({\bf r}_1,{\bf r}_2,i\omega)-
\chi^0_p({\bf r}_1,{\bf r}_2,i\omega)}
{|{\bf r}_1-{\bf r}_2|}\right]
\frac{d{\bf r}_1d{\bf r}_2 d\omega}{2\pi}.\eeq
The variational derivative $\delta\chi^0_p/\delta n_p^l$
has the simple functional form
\beq
\frac{\delta\chi^0_p({\bf r}_1,{\bf r}_2,\omega)}
{\delta n_p^{l}}
=\left[G^p({\bf r}_1,{\bf r}_2,\omega
+\varepsilon_p^{l})
+G^p({\bf r}_1,{\bf r}_2,-\omega
+\varepsilon_p^{l})\right]
\phi^{l*}_{p}({\bf r}_1)
\phi^l_{p}({\bf r}_2),\eeq
where $G^p({\bf r}_1,{\bf r}_2,\omega)$
is the single-particle Green's function for the system of $Z$
noninteracting protons moving in a self-consistent single-particle
nuclear potential. The functional derivative
$\delta\chi_{pp}/\delta n_p^l$ is determined by
the matrix equation
\beq
\frac{\delta\chi_{lm}}{\delta n_p^l}=
\frac{\delta\chi^0_{l}}{\delta n_p^l}\delta_{lm}
+\sum_{k}\left[\frac{\delta\chi^0_{l}}{\delta n_p^l}R_{lk}
\chi_{km}+\chi^0_{l}\frac{\delta R_{lk}}{\delta n_p^l}
\chi_{km}+\chi^0_{l} R_{lk}\frac{\delta\chi_{km}}
{\delta n_p^l}\right].\eeq
which can be derived
by directly varying Eq. (6). Integration with respect to spatial
coordinates is implied in Eq. (21) in just the same way as in Eq.
(6). As can be seen from Eq. (9), the effective interaction
$R_{\alpha\beta}$ is
determined by the form of the functional $F_0$. In order to
simplify the calculations, we took, however, the interaction
$R_{\alpha\beta}$ in the separable representation
\cite{bsja},
\beq R_{lk}({\bf r}_1,{\bf r}_2)=\lambda\frac{d V_l(r_1)}{dr}
\frac{dV_k(r_2)}{dr}\delta(\Omega_1-\Omega_2),\eeq
where $V_l(r)$
is the self-consistent single-particle proton (neutron) potential.
The value of the parameter $\lambda$ is chosen in such a way that the
dipole response has a pole at $\omega=0$. The above separable form of the
effective (residual) interaction is extensively used and, as was
shown in \cite{rs,bm}, provides a good description of collective
nuclear excitations. The value calculated here for the shift
$\Delta\varepsilon_p^l$ of
single-particle proton levels located near the Fermi level is
$(0.2-0.4)$ MeV both for medium-mass and heavy nuclei, in overall
agreement with the value of the Nolen--Schiffer anomaly. It is
worthwhile to verify these results by using the simple LDA
expressions (14) and (15) for $F^{corr}[\rho_p]$. The shift
$\Delta\varepsilon_p^l$ then assumes the form
\beq
\Delta\varepsilon_p^l=\int\frac{\delta F_c^{corr}[\rho_p]}
{\delta\rho_p}|\phi_p^l({\bf r})|^2 d{\bf r}.\eeq
It is convenient to approximate the single-particle
density $\rho_p(r)$ by the Fermi distribution
\beq \rho_p(r,R)=\frac{\rho_0}{1+\exp((r-R)/a)}.\eeq
where $R$ is the nuclear radius and the diffuseness
parameter $a$ is taken to be 0.6 fm. In this case, the
functional $F_c^{corr}[\rho_p]$ given by Eq. (14) reduces to a function
$F_c^{corr}(R)$ of the nuclear radius. An addition of a proton
increases the radius $R$ by the quantity $\Delta R$ that is
determined from the normalization condition
$$\int\left[\rho_p(r,R+\Delta R)-\rho_p(r,R)\right]d{\bf r}=1.$$
The CCE-induced shift of a proton level occurring near
the Fermi surface is given by
$$\Delta\varepsilon_p=F_c^{corr}(R+\Delta R)-F_c^{corr}(R).$$
After simple transformations of the relevant integrals of
the Fermi functions, we find that, for medium-mass and
heavy nuclei, the sought shift of a proton level near the
Fermi level is
\beq\Delta\varepsilon_p\simeq
\frac{D a}{2R}\simeq (0.3-0.4) {\mathrm MeV}.\eeq
Since corrections of order $D(a/R)^3$ were discarded, the
result in (25) slightly overestimates the shift.

The calculated mass differences between mirror
nuclei are quoted in the table. In this calculation, the
functional $F_0$ was constructed on the basis of the $SIII$
interaction \cite{vb}, while the functionals $F_c$ and $F_c^{corr}$
were, respectively, taken in the form (10) and specified by Eqs. (14)
and (15).\\ \bigskip

\noindent
Mass differences between mirror nuclei (in MeV)\\
\begin{tabular}{||l|r|r|r|r|r||} \hline
Nuclei & $^{15}$O  -- $^{15}$N  & $^{17}$F  -- $^{17}$O  &
$^{39}$Ca -- $^{39}$K  & $^{41}$Sc -- $^{41}$Ca
& $^{48}$Ni -- $^{48}$Ca\\ \hline
Theory  & 3.48  & 3.56  & 7.23  & 7.24
& 66.70 \\ \hline
Experiment \cite{n1,bab} & 3.54  & 3.54  & 7.30
& 7.28  & 67.06 \\ \hline \end{tabular}\\

\bigskip
\noindent
The table demonstrates that the mass differences between
mirror nuclei are closely reproduced. The remaining disagreement can
be used to determine the coupling constants for forces violating the
charge symmetry of nuclear interaction. Thus,
traditional nuclear physics can be advantageous in calculating
the constants concerning elementary-particle physics.
Equation (25) shows that the shift $\Delta\varepsilon_p$ of a
single-particle level approximately compensates for the level
shift caused by exchange Coulomb interaction. It
should be recalled that the exchange Coulomb
interaction reduces the energy of a single-particle
level, as can be seen from Eq. (4).
Therefore, the total shift of a
single-particle level due to Coulomb interaction can be
derived by taking into account solely the direct
Coulomb interaction---that is, by retaining only the term
$F_c^H$ in Eq. (10). A similar procedure was postulated in
\cite{fa,bab}. It should be noted that the recipe assuming
mutual cancellation of the exchange Coulomb
interaction and the CCE does not lead to a noticeable value of
the coefficient $a_2^{\prime}$ in Eq. (17) (recall this
coefficient determines the isovector component of the surface
tension) because the contribution from $F_c^H$ and $F_c^F$
[see Eq. (4)] to the surface tension is small. On the contrary,
our calculations indicate that the coefficient $a_2^{\prime}$ is not
small. It also follows from Eq. (25) that, upon taking the
CCE into account, the last filled single-particle proton
level is shifted upward by 0.3 to 0.4 MeV. As a result, the
maximal charge $Z$ of a nucleus occurring near the proton
drip line decreases by 2 or 3 units, as follows from \cite{bm1}.
Therefore, we conclude that, in calculating the proton
drip line, it is important to take the CCE into account.

\section{Effective mass}

Let us consider the change $\Delta M$ that Coulomb  interaction
induces in the effective proton mass $M^*$. In the case of uniform
nuclear matter, the single-particle spectrum depends on the
momentum $p$, while the effective mass, as follows from Eq. (18), is
determined by the expression
\beq
\frac{1}{M^*}=\frac{1}{p_F}
\frac{d\varepsilon_p(p)}{dp}|_{p=p_F},
\eeq
where $p_F$ is the Fermi momentum. In order to calculate
$\Delta M$, we use Eq. (26), replacing $\varepsilon_p(p)$ by the
shift of the single-particle spectrum due to Coulomb interaction.
We have
\beq
\frac{\Delta M}{M^*(M^*+\Delta M)}=
\frac{e^2}{p_F}
\frac{d}{dp}\frac{\delta}{\delta n_p}
\int\left[\frac{\chi_{pp}(q,i\omega) +2\pi\rho_p
\delta(\omega)}{q^2}\right]\frac{dqd\omega}{(2\pi)^3}.
\eeq
Evaluating variations in Eq. (27) and considering that,
in the case of uniform matter, Eqs. (6) and (21) reduce
to algebraic equations, we obtain
\cite{bsja,ks},
\beq \frac{\Delta M}{M^*(M^*+\Delta M)}=\frac{e^2 d}{p_F dp}\int
\frac{\delta\chi_0(q,i\omega)}{\delta n_p}\frac{1}
{(1-R_{+}(q,i\omega)\chi_0(q,i\omega))^2}
\frac{d{\bf q}d\omega}{q^2(2\pi)^3}.\eeq
where $M^*$ is the effective proton mass in the absence of
Coulomb interaction and $R_{+}=(R_{pp}+R_{pn})/2$ is the
effective interaction. The derivative $d/dp$  was taken at the
point $p=p_F$. Let us consider the effective-mass variation
$\Delta M$ for the case where the system under consideration is
characterized by the compressibility parameter $\kappa$,
$1/\kappa=\rho^2d^2E/d\rho^2$,
tending to infinity. This state is similar to states existing
in the surface region of nuclei.  This consideration will enable us
to draw qualitative conclusions on the behavior of the effective mass
in a nucleus. It can be shown that \cite{ks}
\beq \frac{d}{dp}\frac{\delta}{\delta n_p}
\chi_0(q,\omega)|_{p\rightarrow p_F}
=-\frac{4\pi}{p_F^2}\delta(p_F-|{\bf p+q}|) \delta(\omega){\bf
p}({\bf p+q}). \eeq
Substituting (29) into (28) and retaining only the
leading term, we obtain
\beq \frac{1}{M^*+\Delta M}=\frac{1}{M^*} +\frac{e^2}{2\pi
p_F}\int^1_{-1} \frac{x\,dx} {(1-x)[1-R(q(x),0)\chi_0(q(x),0)]^2}.
\eeq
where $q(x)=p_F\sqrt{2(1-x)}$. At the point where the
compressibility diverges, the denominator
$(1-R_{+}\chi_0)^2$ of the
integrand in Eq. (30) vanishes at $x=1\,(q=0)$. The
relevant integral, which is positive, diverges, which results
in the vanishing of the effective mass---that is,
$M^*+\Delta M\to 0$. This result demonstrates that, in equilibrium
nuclear systems, the CCE noticeably affects the effective mass,
the decrease $\Delta M$ in the effective proton mass
being dependent on the degree to which the density of
the system,$\rho$, deviates from its equilibrium value $\rho_0$.
In a nucleus, the divergence of the integral in Eq. (30) --- it could
be due to the decrease in the density near the surface of the nucleus
--- gives way to a finite value. As a result, the effective proton
mass decreases in relation to $M^*$ and becomes smaller than the
effective neutron mass.  It can be said that the effective nucleon
mass acquires an isovector component \cite{bsja}. This possibility
was discussed in \cite{bab}. A theoretical validation of this effect
has been given here.

\section{ Conclusion}

The basic results of the present study can be summarized as follows.
A consistent scheme for constructing the density functional
determining the Coulomb energy of a nucleus has been developed to the
first order in $e^2$.  Using this functional, we have calculated the
single-particle spectra of nuclei and the systematic shift of these
spectra that is induced by the CCE. It has been shown that the
Nolen--Schiffer anomaly is removed to a considerable extent by this
systematic shift of the single-particle spectrum. Owing to the same
mechanism, the proton drip line undergoes a shift of 2 or 3 units
toward smaller values of the charge $Z$ of a nucleus
occurring near this line. It has been shown that the CCE
must be taken into account in calculating the effective
nucleon mass. This opens the possibility of estimating
the coupling constants for forces violating charge
symmetry. The contribution of these forces must be treated
in the same manner as this has been done for the
Coulomb interaction. It can be expected that the proposed
procedure will make it possible to construct a density
functional applicable to describing various properties
of nuclei occurring both in the valley of stability and
beyond it.

\section{Ackowledgments}

I am grateful to A. Bulgac for stimulating discussions on the
problems analyzed in this study. This work was supported in part by
INTAS-OPEN-97-603.

\end{document}